\documentclass[preprint,onecolumn,nofootinbib]{revtex4}
\usepackage[colorlinks=true,linkcolor=blue,urlcolor=blue,filecolor=black,citecolor=red,pdfstartview=FitV,pdftitle={},pdfsubject={},pdfkeywords={},pdfpagemode=None,bookmarksopen=true]{hyperref}
\usepackage{autobreak}
\usepackage{graphicx}
\usepackage{amsmath}
\usepackage{amsfonts}
\usepackage{amssymb,ulem}
\usepackage{color}%
\usepackage{tikz}
\usepackage{dcolumn}
\usepackage{enumerate}
\usepackage{xcolor}
\usepackage{tabularray}

\definecolor{lightblue}{rgb}{0.8, 0.9, 1}
\definecolor{lightred}{rgb}{1, 0.9, 0.8}
\definecolor{lightgreen}{rgb}{0.8, 1, 0.8}

\allowdisplaybreaks

\begin{document}
\title{
	Mixed-State Entanglement and Transport in Einstein-Maxwell-Axion-Horndeski Theory
}

\author{Mu-Jing Li $^{1}$}
\email{mujing.phy@gmail.com}
\author{Chong-Ye Chen $^{1}$}
\email{cychen@stu2022.jnu.edu.cn}
\author{Chao Niu $^{1}$}
\email{niuchaophy@gmail.com}
\author{Cheng-Yong Zhang $^{1}$}
\email{zhangcy@email.jnu.edu.cn}
\author{Peng Liu $^{1}$}
\email{phylp@email.jnu.edu.cn}
\thanks{corresponding author}
\affiliation{
	$^1$ Department of Physics and Siyuan Laboratory, Jinan University, Guangzhou 510632, China
}

\begin{abstract}

	We present a comprehensive study exploring the relationship between transport properties and measures of quantum entanglement in the Einstein-Maxwell-Axion-Horndeski theory. By using holographic duality, we study the entanglement measures, holographic entanglement entropy (HEE) and entanglement wedge cross-section (EWCS), and transport coefficients, for this model and analyze their dependence on free parameters which we classify into action parameter, observable parameters and axion factor. We find contrasting behaviors between HEE and EWCS with respect to observable parameters (charge and temperature), and the axion factor, indicating that they capture different types of quantum correlations. We also find that HEE exhibits positive correlation with both charge and thermal excitations, whereas EWCS exhibits a negative correlation with charge-related conductivities and thermal fluctuations. Furthermore, we find that the Horndenski coupling term, as the modification to standard gravity theory, does not change the qualitative behaviors of the conductivities and the entanglement measures.
	
\end{abstract}
\maketitle
\tableofcontents

\section{INTRODUCTION}

Quantum entanglement, a distinguishing property of quantum systems, plays a crucial role in characterizing quantum phase transitions and the emergence of spacetime \cite{Amico:2007ag,Laflorencie:2016, Horodecki:2009zz, Osterloh:2002, Levin:2006zz, Kitaev:2005dm, Ryu:2006bv, Hubeny:2007xt}. However, calculating quantities related to quantum entanglement poses challenges due to the exponential increase in degrees of freedom and the large Hilbert space dimension in quantum systems \cite{Amico:2007ag, Swingle:2012, Beadle:2008, Islam:2015}.

The AdS/CFT duality offers a powerful tool for studying strongly coupled systems \cite{Maldacena:1997re, Gubser:1998bc, Hubeny:2007xt, Lewkowycz:2013nqa}. One notable application of holographic duality is the holographic entanglement entropy (HEE) introduced by Ryu and Takayanagi \cite{Ryu:2006bv}. HEE relates the minimum surface area of a sub-region in the dual spacetime to the subregion's entanglement entropy \cite{Ryu:2006bv, Ryu:2006ef, Nishioka:2006gr, Klebanov:2007ws, Pakman:2008ui}. This approach simplifies the calculation of quantum information quantities and has found applications in condensed matter theory and QCD \cite{Zhang:2016rcm, Zeng:2016fsb, Ling:2014laa, Ling:2014bda, Hartnoll:2016apf,Jokela:2023lvr}.

One limitation of HEE is its applicability only to pure states, making it unsuitable for measuring the entanglement of mixed states, which are more common in systems at finite temperature \cite{Terhal:2002zz, Vidal:2002zz, Plenio:2005cwa, Calabrese:2014yza, Caputa:2018xuf, Liu:2019qje}. To address this, new measures such as the entanglement of purification and logarithmic negativity have been proposed \cite{Terhal:2002zz, Vidal:2002zz, Plenio:2005cwa, Calabrese:2014yza, Caputa:2018xuf, Liu:2019qje}. The minimum cross-section of two minimum surfaces of subregions, known as the entanglement wedge cross section (EWCS), provides a geometric correspondence for these mixed state entanglement measures \cite{Takayanagi:2017knl, Nguyen:2017yqw, Kudler-Flam:2018qjo, Kusuki:2019zsp, Gong:2020pse, Liu:2021stu, Zhang:2021edm, Chen:2021bjt, Tamaoka:2018ned, BabaeiVelni:2019pkw,Cheng:2021hbw}.

Recently, it is found in condensed matter theory that entanglement can serve as a distinct and novel tool to detect both thermal and quantum phase transitions \cite{Amico:2007ag,Laflorencie:2016,Liu:2020blk}. Quantum phase transitions occur at absolute zero temperature due to quantum fluctuations, whereas thermal phase transitions result from thermal fluctuations. The transport properties of a system, such as electrical and thermal conductivity, often provide insight into its underlying physics. Notably, many quantum and thermal phase transitions demonstrate unique signatures during these transport measurement processes, enabling us to identify the nature of the transitions. In the holographic duality theory, entanglement has also been generally recognized as a novel diagnostic tool for identifying both thermal and quantum phase transitions. Specifically, it has been established that during holographic quantum and thermal phase transitions, the entanglement measures of the system proves crucial in determining the nature of the transition \cite{Baggioli:2017ojd,Ling:2016wyr,Ling:2015dma,Yang:2023wuw}.

While the gravity models without modification terms have been extensively studied, the exploration of modified gravity models, which could exhibit more interesting phenomena in entanglement, is still lacking. Modified gravity theories aim to solve issues from cosmological observations or gravity renormalization by incorporating correction factors to general relativity \cite{Sotiriou:2008rp, Nojiri:2010wj, Buchdahl:1970ynr, Nojiri:2017ncd, Capozziello:2007ec, Panpanich:2018cxo, Stelle:1976gc, Zumalacarregui:2013pma}. One such theory is higher derivative gravity, which includes high-order derivatives of the Riemann curvature tensor \cite{Lovelock:1971yv, Boulware:1985wk, Zwiebach:1985uq, Faulkner:2006ub}. However, the complexity of solving higher-order partial differential equations involved in determining the entanglement measures hinders the exploration of modified models \cite{Dong:2013qoa}.

To address this challenge and investigate the relationship between transport properties and entanglement properties in modified gravity models, we adopt the Einstein-Maxwell-Axion-Horndeski model. The Horndeski model, which is the most general scalar-tensor theory involving a non-minimal coupling between a scalar field and the Einstein tensor, has gained attention as a cosmological model for dark energy and inflation \cite{Horndeski:1974wa, Anabalon:2013oea, Cisterna:2014nua, BeltranJimenez:2013btb, Kobayashi:2014wsa, Liu:2018hzo, Kobayashi:2019hrl, Kase:2018aps, Kobayashi:2011nu, Nicolis:2008in, Feng:2015wvb, Kuang:2016edj, Jiang:2017imk, Baggioli:2017ojd, Geng:2017nwv, Feng:2015oea}. This model is particularly interesting for several reasons. First, the Horndeski model is characterized by a non-minimal coupling between a scalar field and the Einstein tensor, which can significantly affect both the transport properties and the entanglement entropy prescription. This makes it a good candidate for exploring how modifications to gravity can impact these properties. Second, the inclusion of axions breaks the translation symmetry of the system, leading to finite DC conductivities. This feature makes the system more realistic and allows us to examine the transport properties in greater detail. Third, the analytical solution of this system enables us to calculate the DC conductivities analytically, and to work out entanglement-related quantities more easily. By starting with the Horndeski model with axion fields as our first step towards a more general study of transport properties and entanglement in the presence of modifications, we can gain valuable insights into how these phenomena are affected by non-minimal couplings and broken symmetries. 

In this Einstein-Maxwell-Axion-Horndeski model, we adopt the entanglement entropy prescription of Dong \cite{Dong:2013qoa} due to the presence of modification, i.e., the Horndenski coupling. This prescription accounts for the effects of the Horndenski coupling and simplifies the entanglement calculations by involving only ordinary second-order equations of motion. Moreover, we extend the prescription to the EWCS by adjusting the definition of the minimum cross-section based on the modified entanglement entropy prescription. This allows us to explore the entanglement properties in the context of the Horndeski model with axions.

In the subsequent sections, we present the Horndeski spacetime model and introduce holographic information-related quantities, along with the calculation of the equation of motion and holographic thermodynamic conductivities (Section \ref{sec:alg}). We study HEE and EWCS and discuss their relation to transport properties (Sections \ref{sec:HEE}, \ref{sec:eop_phenomena} and \ref{sec:transandent}). Finally, we conclude with a summary in Section \ref{sec:discuss}.

\section{Einstein-Maxwell-Axion-Horndeski Model and Holographic Quantum Information}\label{sec:alg}

In this section, we start by introducing Horndenski gravity theories with axions and explaining their theoretical framework. Then, we delve into the concepts of HEE and EWCS, emphasizing their significance in quantum information theory. Finally, we discuss the computation of these Holographic Information-related Quantities, including the methods and techniques used to calculate them in gravity theories.

\subsection{Einstein-Maxwell-Axion-Horndeski gravity model}\label{sec:EMAH}

The four-dimensional Einstein-Maxwell-Horndeski gravity model with two free axions $\phi_{i}$ can be described by the following Lagrangian density,
\begin{equation} \label{Ld}
	\mathcal{L}=\sqrt{g}\left[\kappa\left(R-2 \Lambda-\frac{1}{4} F^{2}\right)-\frac{1}{2}\left(\zeta g^{\mu \nu}-\gamma G^{\mu \nu}\right) \partial_{\mu} \chi \partial_{\nu} \chi-\frac{1}{2} \sum_{i=1}^{2}\left(\partial \phi_{i}\right)^{2}\right], 
\end{equation}
where $\kappa$ is a constant that we set to 1 for concreteness, and $\Lambda$ is the cosmological constant, which acts as a variable in this model. In addition to the usual gravitational terms, this model includes a scalar field $\chi$, with the derivative of the scaler filed $\chi$ coupled to the Einstein tensor $G_{ab}$. $\zeta$ is the constant governing the magnitude of kinematic energy of the scalar field, and $\gamma$, is the Horndeski coupling. The constants $\zeta$ and $\gamma$ are related by,
\begin{equation}\label{eq:zetagamma}
	\zeta =3g^{2}\gamma,
\end{equation}
where $g=\frac{1}{l}$ is the inverse of the AdS radius. 

It may seem that to turn off the coupling between the scalar field and the gravity, one simply needs to set $\gamma$ and $\zeta$ to zero, returning the action to that of the AdS-RN-Axion theory. However, even in this limit, the modification does not vanish entirely. According to \eqref{eq:alflields}, a finite term will still remain due to the $\chi$ term with a $1/\gamma$ dependence. In other words, this modification will always persist and cannot be switched off. This is the characteristic of the Horndeski model.

The charged Horndeski planar black hole solution in asymptotic AdS spacetime is given by \cite{Liu:2018hzo,Feng:2018sqm,Kuang:2016edj}, 
\begin{equation}\label{metric}
	d s^{2}= -h(r) d t^{2}+\frac{d r^{2}}{f(r)}+r^{2} \sum_{i=1}^{2} d x^{i} d x^{i}, \quad A=a(r) d t, \quad \chi=\chi(r), \quad \phi_{1,2}=k x_{1,2},
\end{equation}
where
\begin{equation}\label{eq:alflields}
	\begin{aligned}
		h=             & g^{2} r^{2}-\frac{2 \kappa k^{2}}{\beta \gamma+4 \kappa}-\frac{m_{0}}{r}+\frac{\kappa\left(3 g^{2} q^{2}(\beta \gamma+4 \kappa)-\kappa k^{4}\right)}{3 g^{2} r^{2}(\beta \gamma+4 \kappa)^{2}} \\
		               & -\frac{\kappa^{2} q^{4}}{60 g^{2} r^{6}(\beta \gamma+4 \kappa)^{2}}-\frac{\kappa^{2} k^{2} q^{2}}{9 g^{2} r^{4}(\beta \gamma+4 \kappa)^{2}},                                                   \\
		f=             & \frac{36 g^{4} r^{8}(\beta \gamma+4 \kappa)^{2}}{\left(6 g^{2} r^{4}(\beta \gamma+4 \kappa)-\kappa\left(q^{2}+2 k^{2} r^{2}\right)\right)^{2}} h,                                              \\
		\chi^{\prime}= & \sqrt{\frac{6 \beta \gamma g^{2} r^{4}-\kappa\left(q^{2}+2 k^{2} r^{2}\right)}{6 \gamma g^{2} r^{4}} \frac{1}{f}},                                                                             \\
		a=             & a_{0}-\frac{q}{r}+\frac{q \kappa k^{2}}{9 g^{2} r^{3}(\beta \gamma+4 \kappa)}+\frac{\kappa q^{3}}{30 g^{2} r^{5}(\beta \gamma+4 \kappa)}.
	\end{aligned}
\end{equation}
In these equations, $\beta$ is a constant related to the scalar field $\chi$, and $q$ is charge of the black hole. The Hawking temperature and the cosmological constant are determined by,
\begin{equation}\label{tem1}
	T=\frac{6 g^{2} r_{h}^{4}(\beta \gamma+4 \kappa)-\kappa q^{2}-2 r_{h}^{2} \kappa k^{2}}{8 \pi r_{h}^{3}(\beta \gamma+4 \kappa)}, \quad \Lambda=-\frac{3 g^{2}(\beta \gamma+2 \kappa)}{2 \kappa}.
\end{equation}
Here, $r_{h}$ is the radius of the horizon of black hole. In this article, we examine the adjustable parameters $(\zeta,\gamma,\beta,q,T,k)$ and group them into three categories:
\begin{enumerate}
	\item Action parameters: These are the parameters $\zeta$, $\gamma$, and $\beta$, which determine the action or theory of the system.
	      
	\item Observable parameters: These include the parameters $T$ and $q$, which are related to observable quantities, the temperature and the charge of the dual system.
	      
	\item Axion factor: The parameter $k$, representing the axion factor, plays a crucial role in the transport properties.
\end{enumerate}

To study the holographic dual of a strongly coupled field theory in Minkowski spacetime, it is necessary to redefine the coordinates as $(t \to \frac{1}{g^2} t, x \to \frac{1}{g} x, y \to \frac{1}{g}y)$ to ensure that the new metric of the asymptotic AdS spacetime takes the standard form at the boundary: $ds^2 =  -r^2 dt^2 + \frac{dr^2 }{r^2} +r^{2} d x^2+r^{2} d y^2 $. Additionally, to simplify the calculations of the EWCS, a change of variable from $r$ to $z$ ($r \to \frac{r_{h}}{z}$) is performed. In this new coordinate system, the black hole horizon is located at $z=1$, while the asymptotic boundary of the spacetime is at $z=0$.

The thermodynamics of AdS black holes in Horndeski theory has been extensively studied in previous works \cite{Feng:2015oea, Feng:2015wvb}. These studies used the Wald formula \eqref{Wald} to analyze the thermodynamic properties of black holes \cite{Wald:1993nt}. The Wald entropy formula \eqref{Wald} is commonly used to analyze the thermodynamic properties of black holes \cite{Wald:1993nt}. In this Horndenski theory, the entropy density $s$ is given by:
\begin{equation}\label{waldentro}
	s=\frac{16 \pi r_{h}(\beta \gamma+4 \kappa)}{3 g^{2}} T.
\end{equation}

In addition to thermodynamics, various transport coefficients, known as conductivities, play an important role in understanding the response of a system to external electric and thermal perturbation. The direct current (DC) thermoelectric conductivities are commonly investigated, which involve constructing a radially conserved current linking the physical properties at the boundary to information present on the black hole horizon \cite{Iqbal:2008by,Blake:2013bqa,Donos:2014uba,Donos:2014cya,Liu:2017kml,Liu:2018hzo}. This can be achieved through perturbating the system, allowing the derivation of radially conserved electric and heat currents on the horizon \cite{Baggioli:2017ojd,Liu:2018hzo}. Based on these calculations, the DC conductivity matrix reads,
\begin{equation}\label{conduct}
	\begin{pmatrix}
		\sigma_{\mathrm{DC}}       & \alpha_{\mathrm{DC}}        \\
		\bar{\alpha}_{\mathrm{DC}} & \bar{\kappa}_{\mathrm{DC}}
	\end{pmatrix}
	=
	\begin{pmatrix}
		\kappa\left(1+\frac{q^{2}}{r_{h}^{2} k^{2}}\right)                 & \frac{4 \pi^{2} q(\beta \gamma + 4 \kappa)}{3 g^{2} r_{h} k^{2}} T         \\
		\frac{4 \pi^{2} q(\beta \gamma + 4 \kappa)}{3 g^{2} r_{h} k^{2}} T & \frac{16 \pi^{4}(\beta \gamma + 4 \kappa)^{2}}{9 \kappa g^{4} k^{2}} T^{3}
	\end{pmatrix},
\end{equation}
where $\sigma_{\mathrm{DC}}$ and $\bar{\kappa}_{\mathrm{DC}}$ are the DC electric and thermal conductivities, respectively, while $\alpha_{\mathrm{DC}}$ and $\bar{\alpha}_{\mathrm{DC}}$ are the DC thermoelectric conductivities. With these analytical expressions for the DC thermoelectric conductivities, we can easily determine how these quantities depend on the action Parameters ($\zeta$, $\gamma$, $\beta$), observable parameters ($q$, $T$), and axion Factor $k$. The relationships between the DC conductivities and parameters are summarized in the first three rows of TABLE \ref{tab:combined}.

Remind that, according to \cite{Donos:2014cya}, in the AdS-RN-Axion model,where the modification terms vanishes, the DC conductivities are given by,
\begin{equation}\label{conduct_axion}
	\begin{pmatrix}
		\sigma_{\mathrm{DC}}       & \alpha_{\mathrm{DC}}        \\
		\bar{\alpha}_{\mathrm{DC}} & \bar{\kappa}_{\mathrm{DC}}
	\end{pmatrix}
	=
	\begin{pmatrix}
		\frac{q^{2}}{k^{2}r_{h}^{2}}+1 & \frac{4 \pi q}{k^{2}}                 \\
		\frac{4 \pi q}{k^{2}}          & \frac{16 \pi^{2}}{k^{2}} r_{h}^{2} T
	\end{pmatrix},
\end{equation}
Therefore, it is deduced that, the relationship between the DC conductivities and parameters are similar to the case of the current model.

In light of the preceding overview of the EMAH model, we will now proceed to introduce the holographic quantum information measures.

\subsection{Holographic Entanglement Entropy and Entanglement Wedge Minimum Cross-section}

The von Neumann entropy is a fundamental concept in quantum mechanics that quantifies the entanglement between a subsystem, denoted as $A$, and its complement in a pure state $|\psi\rangle$ \cite{Nielsen:2002}. It serves as the quantum counterpart of classical information entropy and is commonly referred to as the entanglement entropy (EE). The EE of subsystem $A$ can be computed using the following formula,
\begin{equation}
	S_{A}(|\psi\rangle)=-\operatorname{Tr_{A}}\left[\rho_{A} \log \rho_{A}\right],
\end{equation}
where $\rho_{A}$ represents the reduced density matrix of subsystem $A$, obtained by tracing out subsystem $B$ from the total density matrix $\rho_{A}=\operatorname{Tr}_{B}(|\psi\rangle\langle\psi|)$.

The holographic duality of entanglement entropy, known as holographic entanglement entropy (HEE), was initially proposed by Ryu and Takayanagi in 2006 based on the holographic principle, marking a significant breakthrough in the field of quantum information \cite{Ryu:2006bv}.
In holography, the entanglement entropy is related to the geometry of a certain region in the gravitational theory. Specifically, the entanglement entropy $ S_A $ of a certain subregion $ A $ in the quantum system is proportional to the area of a minimal surface that spans the boundary of region $ A $ in the gravitational theory,
\begin{equation}
	S_A = \frac{{\text{{Area}}(\gamma_A)}}{{4G_{\text{{N}}}}}
\end{equation}
where $ \gamma_A $ is the minimal surface and $ G_N $ is the Newton's gravitational constant.

The EE measures the amount of entanglement between two subsystems in a pure state. It quantifies the degree to which the state cannot be factored into separate states for each subsystem. For a pure state, the entanglement entropy is non-zero, while for a fully factorizable state (no entanglement), the entropy is zero. However, when dealing with mixed states, which are ensembles of different pure states, the entanglement entropy is not a good diagnostic since it can be non-zero even in the absence of entanglement. To overcome this limitation, one can use the mutual information (MI). The MI quantifies the amount of shared information between two subsystems, regardless of whether it arises from entanglement or from classical correlations. It is defined as the difference between the entanglement entropies of the combined region $ A\cup B $ and the individual regions $ A $ and $ B $:
\begin{equation}
	I(A; B) = S_A + S_B - S_{A \cup B}
\end{equation}
The MI provides a more complete picture of the correlations in a quantum system, as it captures both classical and quantum correlations. In the case of a pure state, the MI is equal to the entanglement entropy, but for mixed states, it can also capture classical correlations that do not involve entanglement. Hence, MI is not a perfect measure of entanglement in mixed states, as it can be dominated by the entanglement entropy.

\begin{figure}[htbp]
	\includegraphics[height=0.3\textwidth]{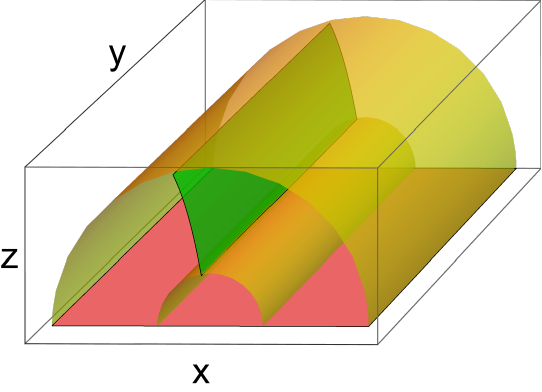}
	\includegraphics[height=0.3\textwidth]{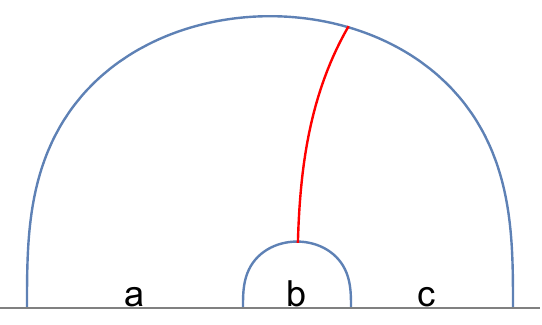}
	\caption{Visualization of the minimal cross-section (green surface on the left plot and red string on the right plot) that cuts through the entanglement wedge.}
	\label{fig:stripe3}
\end{figure}

An alternative technique is the entanglement of purification (EoP), which involves a double minimization procedure to purify the extra degrees of freedom of mixed states \cite{Terhal:2002zz}. 
The holographic EoP has been proposed as proportional to the area of the minimal cross-section of the entanglement wedge (EWCS, see Fig. \ref{fig:stripe3}) \cite{Takayanagi:2017knl}. Later, the EWCS, has been proposed to be dual to other mixed state entanglement measures such as logarithmic negativity \cite{Kusuki:2019zsp,Kudler-Flam:2018qjo,Jain:2020rbb}. 

Building upon the earlier discussion on holographic quantum information measures, we will now delve into their computational methods in the Axion-Horndeski model. In this scenario, the entanglement entropy prescription undergoes substantial modifications, resulting in a more intricate calculation procedure.

\subsection{Computation of the Holographic Information-related Quantities}

Now, we will discuss the computation methods for holographic information-related quantities. Our discussion will start by explaining how to obtain the minimum surfaces for HEE and EWCS. Once we have established this foundation, we will then elaborate on the methods for determining the minimal cross-section within the entanglement wedge.

The entanglement entropy of Horndeski can be evaluated by substituting the Lagrangian density \eqref{Ld} into the entanglement entropy prescription proposed by Dong in \cite{Dong:2013qoa},
\begin{equation} \label{dong}
	\begin{aligned}
		S=\frac{\kappa}{4}\int d^{d}y\sqrt{h} \Big{\{}- \frac{\partial \mathcal{L} }{\partial R_{\mu \nu  \rho\sigma}} \varepsilon_{\mu \rho}  \varepsilon _{\nu \sigma }+ 
		\sum_{\zeta }\left (  \frac{\partial^2 \mathcal{L}}{\partial R_{\mu_{1} \rho_1 \nu_1 \sigma_1} \partial R_{\mu_2 \rho_2 \nu_2 \sigma_2}}\right )_\zeta 
		\frac{2K_{k_1 \rho_1 \sigma_1}K_{k_2 \rho_2 \sigma_2}}{q_\zeta +1} 
		\times \\ 
		\left [ \left ( n_{\mu_1 \mu_2}n_{\nu_1 \nu_2}- \varepsilon _{\mu_1 \mu_2}\varepsilon _{\nu_1 \nu_2} \right ) n^{k_1 k_2}  +
			\left ( n_{\mu_1 \mu_2}\varepsilon _{\nu_1 \nu_2}+   \varepsilon _{\mu_1 \mu_2}n_{\nu_1 \nu_2} \right )  \varepsilon ^{k_1 k_2} \right ] \Big{\}}. 
	\end{aligned}
\end{equation}
The Greek letters are indices of tensors in the $d$-dimensional spacetime. Several key terms are defined as  
\begin{equation} \label{dongsubdef}
	\begin{aligned}
		n_{\mu \nu}           & =n_{\mu}^{(a)} n_{\nu}^{(b)} G_{a b},                       \\
		\varepsilon_{\mu \nu} & =n_{\mu}^{(a)} n_{\nu}^{(b)} \varepsilon_{a b},             \\
		K_{\lambda \mu \nu}   & =n_{\lambda}^{(a)} m_{\mu}^{(i)} m_{\nu}^{(j)} K_{a i j}. 
	\end{aligned}
\end{equation}
Here, $n^{(a)}_{\mu}$ and $m^{(i)}_{\mu}$ are unit vectors perpendicular and transversal to the surface, respectively. The Latin letters ${(a),(b)}$ and ${(i),(j)}$ represent the internal indices in the tangent space. The induced metric on the tangent space is denoted by $G_{ab}$, while $\varepsilon_{ab}$ and $K_{aij}$ represent the Levi-Civita tensor and the extrinsic curvature, respectively.
The coefficient $q_\zeta$ corresponds to an anomaly coefficient associated with each term in the expansion \cite{Dong:2013qoa}.

However, it is important to note that in the case of Horndeski Lagrangian density \eqref{Ld}, only a first-order term of the Riemann tensor is present. Calculations using this Lagrangian density will yield results that are equivalent to the Wald entropy formula, which is given by,
\begin{equation}\label{Wald}
	S=\frac{\kappa}{4}\int d^{d}y\sqrt{h} \left(- \frac{\partial \mathcal{L} }{\partial R_{\mu \rho \nu \sigma}} \varepsilon_{\mu \rho}  \varepsilon _{\nu \sigma } \right ).
\end{equation}
By extracting each curvature-related term from the Lagrangian density, one obtains the following expression for the entanglement entropy \cite{Caceres:2017lbr},
\begin{equation}\label{eom}
	S=\frac{\kappa}{4}\int d^{d}y\sqrt{h} \left [ 1-\frac{\gamma }{4\kappa }
		h_{ij}\chi ^{,i}\chi ^{,j} \right ].
\end{equation}
The $h_{ij}$ represents the induced metric on the surface. We can treat \eqref{eom} as an action and obtain the equation of the minimum surface through the variational principle\footnote{For the definition of the indices in this context, please refer to \cite{Caceres:2017lbr}.} \cite{Caceres:2017lbr},
\begin{equation}\label{eq:form:1}
	\left ( 1-\frac{\gamma G}{4}h^{ij}\chi_{,i} \chi _{,j}  \right ) K_{a}+\frac{\gamma G}{2} K^{ij}_{a}
	\chi_{,i} \chi _{,j}+\frac{\gamma G}{2}\chi _{,a} h^{ij}D_{i} D_{j}\chi =0.
\end{equation}
However, the explicit expression of equation of motion of \eqref{eq:form:1} is quite complicated. 

\begin{figure}[h]
	\centering
	\includegraphics[scale=0.8]{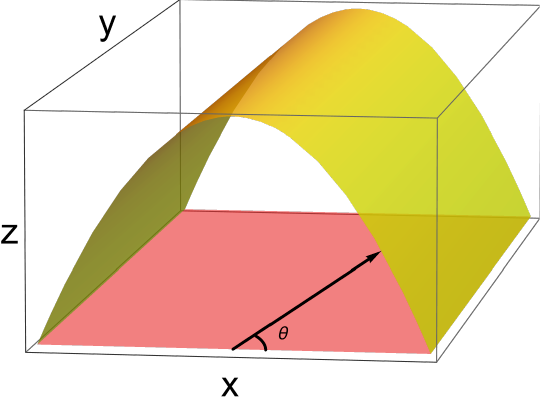}
	\caption{\raggedright A visualization of the minimal surface (yellow) corresponding to an infinite strip (red) along the $y$ axis.}
	\label{fig:stripe}
\end{figure}
To simplify our computations, we follow the computational methods in \cite{Liu:2019qje} and consider an infinite strip along the $y$ axis as the subregion (see Fig. \ref{fig:stripe}). By parameterizing the surface with $\theta$ as $\theta = \arctan(z/x)$, we can derive the equations of motion more easily. To obtain the minimum surface, we need to solve the equations of $z(\theta)$ and $x(\theta)$ within the range of $ \theta \in [0,\pi]$. However, since there is a mirror symmetry about $\theta=\pi/2$, it suffices to work within $\theta \in [0, \pi/2]$.
We treat the integrand in equation \eqref{eom} as a Lagrangian density of the minimum surface. Resultantly, the minimum surface can be obtained by solving the equations of motion from this Lagrangian density,
\begin{equation}\label{ll}
	\mathcal{L}_{\text{HEE}}=\sqrt{g_{xx}(z) x'^2+g_{zz}(z) z'^2} \sqrt{g_{yy}(z) }\left(1-\frac{\gamma z'^2 \chi'(z)^2}{4 \left(g_{xx}(z) x'^2+g_{zz}(z) z'^2\right)g_{yy}(z) }\right).
\end{equation}
Here, $z'$ and $x'$ are derivatives with respect to $\theta$. By taking the variational derivative of \eqref{ll} with respect to $z(\theta)$ and $x(\theta)$ directly, we can derive the equations of motion for $z$ and $x$. Since these equations are highly non-linear ordinary differential equations, solving them can be challenging. One approach is to use the Chebyshev collocation and Newton iteration algorithm. This iterative method is commonly used to solve boundary value problems and involves approximating the solution by a polynomial that satisfies the differential equations at a set of collocation points. The algorithm then updates the polynomial coefficients using Newton's method until a satisfactory solution is obtained. Although this method can be computationally intensive, it has proven to be effective in solving similar types of problems.

After obtaining the minimum surface, the HEE can be calculated through numerical integration using the Lagrangian density \eqref{ll} of the minimum surface. However, the HEE diverges at the AdS boundary ($z=0$), so we need to identify and subtract the diverging term to obtain a finite value. The diverging term is given by,
\begin{equation}\label{eq:heediv}
	\text{div}=\frac{r_h \sqrt{r_h^{2} x^{\prime}(0)^{2}+ z^{\prime}(0)^{2}}}{g z(0)^{2}},
\end{equation}
which needs to be subtracted so as to render a finite value of HEE. 

Next, we elaborate on how to compute EWCS. To determine the EWCS, we first identify the minimum cross section within the entanglement wedge. This involves finding two minimum surfaces and then locating the minimum cross-section that connects them. In order to parametrize the minimum cross-section, we use the variable $z$, which has been verified as a good parametrization in our numerics. The Lagrangian density \eqref{ll} under this parametrization is given by,
\begin{equation}\label{eq,ll2}
	\mathcal{L}_{\text{EWCS}}=\sqrt{g_{xx}(z) x'(z)^2+g_{zz}(z)}\left(1-\frac{\gamma  \chi '(z)^2}{4
	\left(g_{xx}(z)x'(z)^2+g_{zz}(z) \right)}\right).
\end{equation}
The equation of motion for $x(z)$ is obtained by variating \eqref{eq,ll2} with respect to $x(z)$. In our previous work \cite{Liu:2020blk}, we proposed an algorithm to identify the minimal cross-section based on the geometric fact that the minimal cross-section must be perpendicular to the minimum surfaces. This observation significantly speeds up the computation process. However, when considering the modified Lagrangian where the scalar field also takes part in the dual information-related quantities, we need to examine whether the perpendicular condition still holds. Essentially, the perpendicular condition comes from the term $\sqrt{g_{xx}(z) x'(z)^2+g_{zz}(z)}$ and is valid for any form of Lagrangian once this term appears as a factor. Therefore, the algorithm in \cite{Liu:2020blk} still applies to the current situation. Finally, we can work out the area of this minimum cross-section by integrating \eqref{eq,ll2} over $z$.

Besides the computational methods discussed earlier, a primary focus of this paper is on exploring the thermodynamic and transport properties of the Horndenski gravity model with axions and their connections to holographic information-related quantities. In the subsequent section, we provide detailed explanations of these holographic information-related quantities and thermodynamic and transport properties.

\section{The Entanglement-Related Quantities and Transport Properties}\label{sec:entandtrans}

\subsection{Holographic Entanglement Entropy}\label{sec:HEE}

In this section, we will examine the influence of action Parameters ($\zeta$, $\gamma$, $\beta$), observable parameters ($q$, $T$), and axion Factor $k$ on holographic entanglement entropy (HEE). Since the divergence of the HEE \eqref{eq:heediv} is influenced by the parameter $g$, we fix $g=3$ when computing the HEE to enable a valid comparison between different parameters. As a result, \eqref{eq:zetagamma} means that the HEE will exhibit the same behavior with $\gamma$ and with $\zeta$. Therefore, for HEE we will only show HEE vs $\zeta$ and $\beta$.

\begin{figure}
	\centering
	\includegraphics[height=0.3\textwidth]{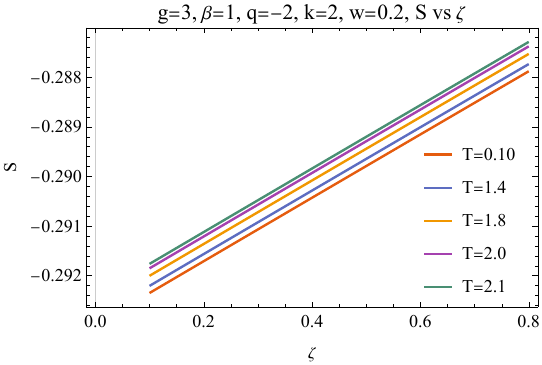}
	\includegraphics[height=0.3\textwidth]{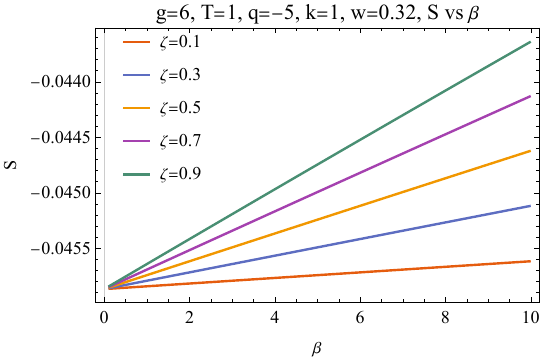}
	\caption{The influence of action Parameters $\zeta$ and $\beta$ on holographic entanglement entropy (HEE). Left: Variation of HEE with different values of $\zeta$. Right: Variation of HEE with different values of $\beta$.}
	\label{fig:action_parameters}
\end{figure}

First, from the action parameter aspect, we find that, HEE increases with $\zeta$ and $\beta$, as shown in Fig. \ref{fig:action_parameters}. In the action, $\zeta$ governs the magnitude of the kinetic energy of the scalar field $\chi$, while $\beta$ is involved in the definition of the scalar field $\chi$ itself. The parameter $\gamma$ quantifies the strength of gravity coupling with the scalar field. The observation that the HEE increases with both $\zeta$ and $\beta$ suggests that enhancing either the magnitude of the kinetic energy or the scalar field itself leads to a greater entanglement of the holographic degrees of freedom.

\begin{figure}
	\centering
	\includegraphics[height=0.3\textwidth]{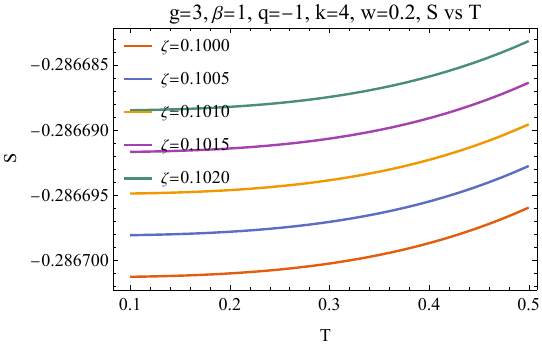}
	\includegraphics[height=0.3\textwidth]{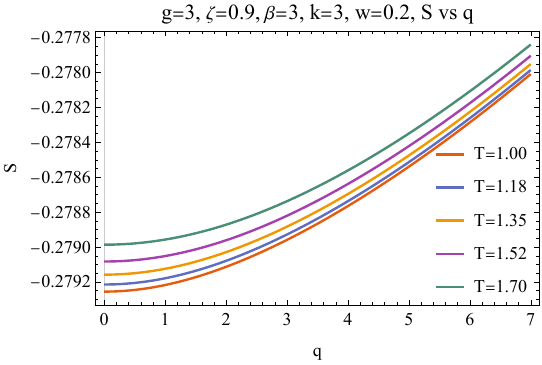}
	\caption{The influence of observable parameters $q$ and $T$ on HEE. Left: Variation of HEE with different values of $q$. Right: Variation of HEE with different values of $T$.}
	\label{fig:observable_parameters}
\end{figure}

Second, when considering the observable parameters, black hole charge $q$ and the Hawking temperature $T$, it has been observed that the HEE monotonically increases with both of these parameters. This observation is depicted in Fig. \ref{fig:observable_parameters}. This behavior can be understood from both the gravity side and the dual CFT side. From the gravity side, the increase in HEE with black hole charge and temperature can be attributed to the growth of the black hole horizon. As the charge $q$ increases, the horizon area expands, leading to an increase in the Bekenstein-Hawking entropy. The HEE, which is proportional to the area of the minimal surface in the bulk theory holographically dual to the boundary CFT, follows a similar trend and thus increases with the charge. Similarly, an increase in temperature leads to a larger black hole horizon and subsequent increase in the HEE. This result is consistent with our expectation that larger black holes have more entanglement entropy. From the dual CFT side, the monotonically increasing HEE with both charge and temperature can be understood by considering the behavior of the dual conformal field theory. The HEE measures the quantum entanglement between degrees of freedom in the CFT. As the charge and temperature increase, the system becomes more excited, leading to an enhanced entanglement between its constituents. This increased entanglement is reflected in the larger value of the HEE.

\begin{figure}
	\centering
	\includegraphics[height=0.3\textwidth]{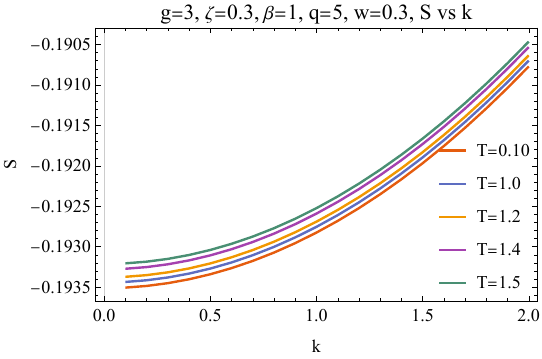}
	\caption{The influence of axion Factor $k$ on holographic entanglement entropy (HEE).}
	\label{fig:axion_and_width}
\end{figure}

The HEE versus the axion factor $k$ is depicted in Fig. \ref{fig:axion_and_width}. As observed, HEE exhibits a monotonically increasing trend with increasing $k$. In the context of holographic duality, $k$ is commonly associated with the transport properties of the system. By breaking the translational symmetry, $k$ allows for the emergence of finite DC conductivities. This implies that as $k$ is increased, the system experiences a smaller DC resistivity, accompanied by an increase in HEE. From the perspective of gravity, it can be argued that the increase in HEE with $k$ is due to the modification in the gravitational background caused by the axion field. Similar results have been observed in \cite{Liu:2020blk} and the mechanisms for this monotonic behavior can be attributed to the relationship between the horizon radius and the parameter $k$. On the dual field theory side, the enhanced EE can be attributed to the extra degrees of freedom aside from the confined charged degrees of freedoms, such as the enhanced lattice vibrations.

Generally, we would expect HEE to exhibit the same pattern as the thermodynamics entropy \eqref{waldentro} when the width of the subregion is large. All these behaviors have been summarized in TABLE \ref{tab:combined}.

\subsection{Entanglement Wedge Cross-section}\label{sec:eop_phenomena}

We will now investigate how varying different parameters affects the EWCS, while keeping the configuration $(a, b, c)$ fixed. It is important to note that the results we obtained are independent of the specific choices of $(a, b, c)$. Therefore, we will only present a few characteristic plots. In the case of EWCS, $g$ is a free parameter, allowing us to consider $\gamma$ and $\zeta$ as two independent variables. Next, we examine the behavior of the EWCS in relation to the action parameters $\zeta,\gamma,\beta$, as well as the observable parameter $q, T$, and axion factor $k$. 

\begin{figure}
	\centering
	\includegraphics[height=0.22\textwidth]{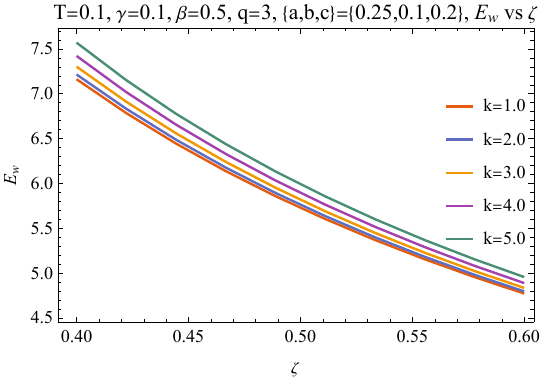}
	\includegraphics[height=0.22\textwidth]{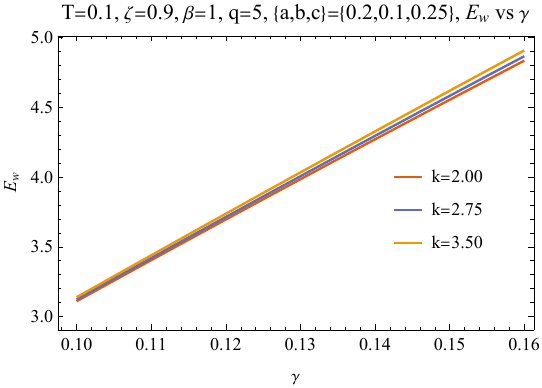}
	\includegraphics[height=0.22\textwidth]{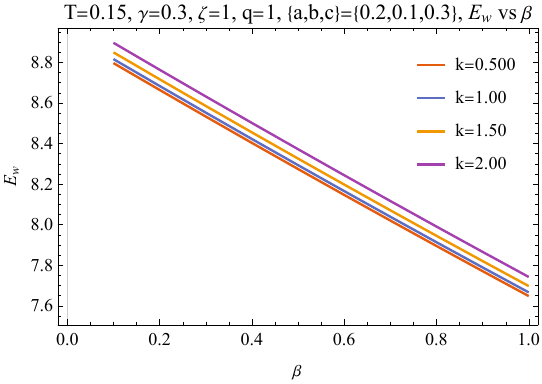}
	\caption{The influence of action Parameters $\zeta,\gamma$ and $\beta$ on the EWCS. Left: Variation of EWCS with different values of $\zeta$. Right: Variation of EWCS with different values of $\gamma$. Right: Variation of EWCS with different values of $\beta$.}
	\label{fig:ewcs_action_parameters}
\end{figure}

Let us start with $\zeta$, which is related to the magnitude of the kinetic energy of the scalar field. We observe that as $\zeta$ increase, the EWCS decreases (see the left plot of Fig. \ref{fig:ewcs_action_parameters}). From the perspective of gravity, this can be understood as a consequence of the scalar field possessing more kinetic energy. With a larger $\zeta$, the gravitational effects induced by the scalar field become stronger. This increased gravitational impact leads to modifications in the spacetime geometry, resulting in a decrease of the EWCS and reducing the entanglement across the entanglement wedge.

Moving on to $\gamma$, which represents the strength of the coupling between gravity and the scalar field, we find that the EWCS increases as $\gamma$ increases (see the middle plot of Fig. \ref{fig:ewcs_action_parameters}). From the gravity side, this can be understood as a consequence of stronger gravitational effects due to the scalar field. As $\gamma$ increases, the coupling between the scalar field and gravity becomes stronger, resulting in enhanced entanglement across the entanglement wedge. Consequently, the EWCS exhibits an increase. 

Lastly, we consider the effect of $\beta$, which is embedded within the definition of the scalar field $\chi$. As $\beta$ increases, we observe a decrease in the EWCS (see the right plot of Fig. \ref{fig:ewcs_action_parameters}). Remind that, HEE exhibits a monotonically increasing behavior. This suggests that HEE exhibit a opposite behavior with $\beta$ when compared with EWCS. This means that, EWCS as a mixed state entanglement measure, captures distinct features of the entanglement from the EE.

\begin{figure}
	\centering
	\includegraphics[height=0.3\textwidth]{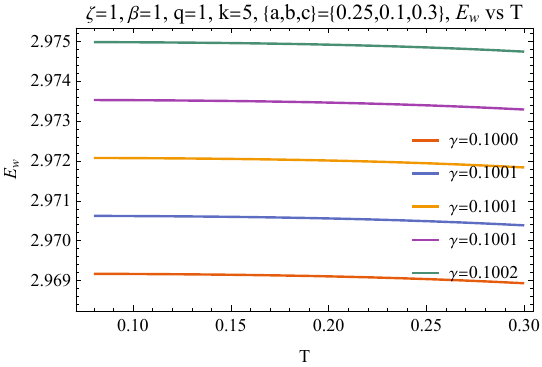}
	\includegraphics[height=0.3\textwidth]{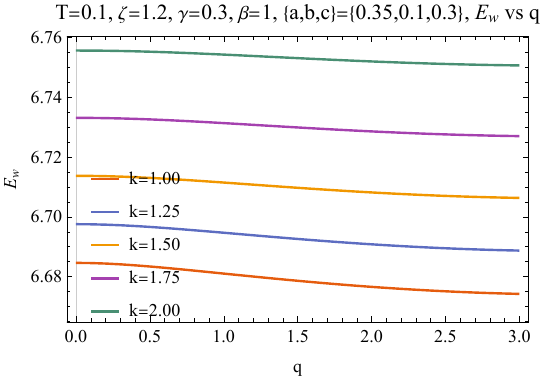}
	\caption{The influence of observable parameters $q$ and $T$ on the EWCS. Left: Variation of EWCS with different values of $q$. Right: Variation of EWCS with different values of $T$.}
	\label{fig:ewcs_observable_parameters}
\end{figure}

Now, we study how the observable parameters $T$ and $q$ affects the EWCS. The EWCS is found to decrease with increasing temperature $T$ and charge $q$, as depicted in Fig. \ref{fig:ewcs_observable_parameters}. This phenomenon is in contrast to the behavior of HEE which typically increases with temperature and charge.

From a gravitational viewpoint, we can gain an intuitive understanding of the EWCS based on the reasoning behind why HEE increases with $ q $ and $ T $. As temperature and charge increases, the black hole horizon expands, causing HEE to resemble thermal entropy more closely. In thermal systems, different subregions should be disentangled, as the total density matrix can be approximately represented by the direct product of each subregion's density matrix. This concept is easily understood from a gravity perspective: when both minimum surfaces approach the horizon, the cross-section becomes smaller due to the narrowing entanglement wedge.

On the dual CFT side, a possible understanding for the decreasing EWCS with increasing temperature and charge can be derived from the behavior of entanglement in strongly coupled systems. At higher temperatures, the thermal fluctuations in the system become more significant. These fluctuations can disrupt or break the entanglement between particles, leading to a decrease in overall entanglement. Increased temperatures generally introduce more randomness and decoherence into the system, which can result in the loss of entanglement. Additionally, the presence of a higher charge may lead to stronger interactions between particles. These interactions can also contribute to decoherence and disrupt the entanglement present in the system. The increased charge density may cause stronger repulsion, which can break or reduce the entanglement between particles. One example where charge density and interactions can affect entanglement is in the context of the Hubbard model, which describes interacting particles on a lattice. In this model, increasing charge density can lead to stronger on-site interactions between particles, which can influence the entanglement properties of the system. 

\begin{figure}
	\centering
	\includegraphics[height=0.3\textwidth]{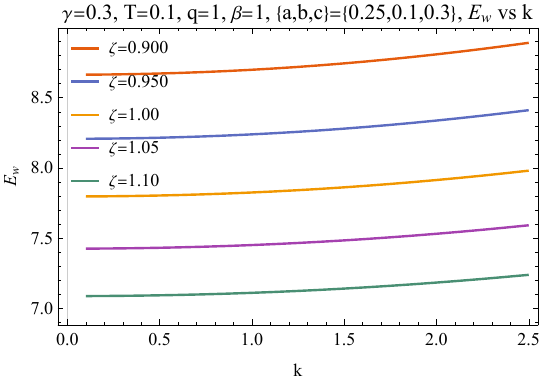}
	\caption{The influence of axion Factor $k$ on EWCS. }
	\label{fig:ewcs_axion_and_width}
\end{figure}

The EWCS versus the axion factor $k$ also depicts a similar trend as the HEE. It exhibits a monotonically increasing behavior with increasing $k$, as can be seen in Fig. \ref{fig:ewcs_axion_and_width}. This observation suggests that mixed state entanglement measures, such as the EWCS, also show an enhanced entanglement when the system possesses a larger axion factor $k$. From the gravity side, the increase in EWCS can be understood as a consequence of the modification in the gravitational background caused by the axion field. The interplay between the axion field and the gravitational dynamics affects the entanglement characteristics of the system, leading to a higher value of EWCS as $k$ increases. This phenomenon has been studied in \cite{Liu:2020blk}, where a similar monotonic behavior of the entanglement measures with $k$ was observed. The specific mechanisms underlying this relationship between the axion factor and the entanglement measures may be attributed to the interplay between the horizon radius and the parameter $k$, as well as other relevant properties of the gravitational background. On the dual field side, the enhanced mixed state entanglement can be attributed to the effects of localization and confinement resulting from the breaking of translational symmetry by the axion field. The breaking of translational symmetry induces a more localized and confined behavior for charged degree of freedom of the quantum states, meanwhile also induces an enhanced extra degrees of freedom such as lattice vibration, resulting in an increased mixed state entanglement.

In the study of modified gravity models, it has been observed that the Horndeski model with axion and the AdS-RN-Axion model (without modification) exhibit similar behaviors in terms of their DC conductivities. This means that the inclusion of the modification term does not significantly alter the DC conductivities in these models. Furthermore, it has also been found that the HEE and EWCS are quantitatively the same for both models\footnote{For the HEE and EWCS study on the pure axion model, see \cite{Huang:2019zph}.}. These findings suggest that when axions are minimally coupled to gravity, the modification term does not have a significant impact on the quantum information, such as the DC conductivities and HEE/EWCS, and their relationships.

\subsection{The Relationship Between the Transport Properties and the Entanglement-Related Quantities}\label{sec:transandent}

\begin{table}
	\centering
	\begin{tblr}{
		colspec      = {|>{\centering\arraybackslash}m{6cm}|*{6}{>{\centering\arraybackslash}m{1.cm}|}},
		column {2-4} = {lightblue},
		column {5-6} = {lightred},
		column {7}   = {lightgreen},
		}
		\hline
		Quantity/Parameter              & $\gamma$     & $\zeta$      & $\beta$      & $T$          & $q$          & $k $         \\ \hline\hline
		$\sigma_{\mathrm{DC}}$          & $\downarrow$ & $\uparrow$   & $\uparrow$   & $\downarrow$ & $\uparrow$   & $\downarrow$ \\ \hline
		$\alpha_{\mathrm{DC}}$          & $\uparrow$   & $\downarrow$ & $\uparrow$   & $\uparrow$   & $\uparrow$   & $\downarrow$ \\ \hline
		$\bar{\kappa}_{\mathrm{DC}}$    & $\uparrow$   & $\downarrow$ & $\uparrow$   & $\uparrow$   & $\downarrow$ & $\downarrow$ \\ \hline\hline
		\textbf{EWCS}                   & $\uparrow$   & $\downarrow$ & $\downarrow$ & $\downarrow$ & $\downarrow$ & $\uparrow$   \\ \hline
		\textbf{HEE}                    & $\downarrow$ & $\downarrow$ & $\uparrow$   & $\uparrow$   & $\uparrow$   & $\uparrow$   \\ \hline
		\textbf{Thermodynamics Entropy} & $\downarrow$ & $\downarrow$ & $\uparrow$   & $\uparrow$   & $\uparrow$   & $\uparrow$   \\ \hline
	\end{tblr}
	\caption{
		The dependence of conductivities, EWCS, HEE, and thermodynamic entropy on variations of different parameters. An upward arrow indicates an increase with the parameter, while a downward arrow indicates a decrease with the parameter.
	}
	\label{tab:combined}
\end{table}

The relationship between transport properties and entanglement related quantities is a topic of great interest in the study of condensed matter physics. Transport properties, such as DC electrical conductivities, play a crucial role in characterizing the behavior of materials, distinguishing between metals and insulators, and probing phase transitions.

Recent research has revealed intriguing connections between transport properties and entanglement properties, particularly in the context of condensed matter systems and holographic frameworks. In certain scenarios, the entanglement properties exhibit extremal behaviors across phase transitions and in strongly correlated systems. These findings suggest that there may be underlying connections between the transport properties and entanglement properties.

In the current Horndeski model with axion, we also systematically analyze the relationship between transport properties and entanglement in this model. To illustrate this more clearly, we present all the results in Table \ref{tab:combined}. In Table \ref{tab:combined}, the rows and columns represent physical quantities and the free parameters of the system, respectively. In the columns, for better comparison, we highlight the action parameter part in blue, observable parameters in orange, and the axion factor in green. In the rows, the first three rows represent the transport coefficients of the system, while the last three rows represent the quantities related to entanglement and the thermodynamic entropy density.

Let us now examine the observable parameters, namely $q$ and $T$. Increasing the value of $q$ leads to an increase in charge-related conductivities ($\sigma_{\text{DC}}$ and $\alpha_{\text{DC}}$), while the thermal conductivities decrease. This can be understood in the context of condensed matter systems, such as insulators, where the primary mode of heat transfer is through lattice vibrations (phonons) rather than free electron movement. In such systems, an increase in charge density does not enhance thermal conductivity. In fact, high charge densities can even decrease thermal conductivity by increasing electron-electron scattering, which hinders the flow of heat. Additionally, increasing $q$ is always associated with higher EE. This is because higher charge densities result in more charged degrees of freedom, leading to increased entanglement as well as the thermal entropy, as explained in previous sections. However, the mixed state entanglement EWCS, decreases with increasing $q$, which can result from the localization induced by strong correlation between electrons. Meanwhile, the thermal conductivities exhibit the opposite trend. This is understandable since enhanced heat transfer tends to disrupt entanglement.

For the temperature $T$, the heat-related DC conductivities ($\bar\kappa_{\text{DC}}$ and $\alpha_{\text{DC}}$) increase with increasing $T$, while $\sigma_{\text{DC}}$ decreases. This behavior can be explained by the fact that higher temperatures facilitate more efficient heat transfer, resulting in larger values of $\bar\kappa_{\text{DC}}$ and $\alpha_{\text{DC}}$. The decreasing behavior of $\sigma_{\text{DC}}$ indicates that the system is in a typical metallic phase. Turning our attention to the entanglement-related quantities, we find that increasing the temperature leads to an increase in the entanglement entropy (HEE) due to the contribution of thermal entropy, while the EWCS decreases due to the disruption of thermal effects.

Based on the above observations, it appears that the HEE is aligned with the thermo-electric conductivity $\alpha_{\text{DC}}$. Which means that HEE detects both the heat and charged degrees of freedom. 
Meanwhile, EWCS is negatively related to the charged degrees of freedom in the sense that, on the $q$ direction, $\sigma_{\text{DC}}$ and $\alpha_{\text{DC}}$ both increases with increasing charge density, while EWCS decreases due to the strong correlation of the charged degrees of freedom.
Also, EWCS is aligned with $\sigma_{\text{DC}}$ on the temperature direction, but the mechanisms are different at that, $\sigma_{\text{DC}}$ decreases as a metallic charge transport behavior, but the thermal effects tends to disrupt the mixed state entanglement, i.e., EWCS. 

Next, let us focus on the axion factor $k$. All three DC conductivities decrease as the value of $k$ increases. This decrease is largely attributed to the process of momentum relaxation, which not only affect momentum but also impacts the flow of charge and energy. Keep in mind that this pattern has been observed in several other models, thereby suggesting that it might be a common occurrence within such axion structures, particularly within those that are minimally coupled. Conversely, as $k$ increases, the entanglement-related quantities such as the HEE, and the EWCS all increase. This seems counterintuitive, as a reduction in the flow of charge — especially when charged degrees of freedom are predominant — tends to significantly decrease entanglement. We present this as evidence that in this model, despite decreasing charge mobility and associated momentum and energy, the charged degrees of freedom does not dominate the system's entanglements. It suggests that the augmentation of other degrees of freedom, such as phonon, takes place, resulting in higher entanglement. 

\section{Discussion}\label{sec:discuss}

This work presents a comprehensive study of holographic entanglement entropy (HEE) and entanglement wedge cross-section (EWCS) in the Horndeski gravity model with axions. Leveraging holographic duality, we investigate the interplay between these measures of quantum entanglement and the transport properties described by DC thermoelectric conductivities. Our analysis reveals several key insights:

\begin{itemize}
	\item The HEE increases with the action parameters $\zeta$, $\beta$, black hole charge $q$, and temperature $T$. This aligns with expectations from black hole thermodynamics, where entropy increases with charge and temperature. The EWCS exhibits the opposite trend, decreasing with $q$ and $T$.
	\item The contrasting behavior of HEE and EWCS suggests that they capture different types of entanglement. HEE appears more closely related to the degrees of freedom associated with charge and thermal excitations. In contrast, EWCS seems more influenced by disruptive effects like thermal fluctuations that can destroy quantum correlations.
	\item All DC conductivities decrease with the axion factor $k$, indicating momentum relaxation effects that hinder charge and energy transport. However, both HEE and EWCS increase with $k$, highlighting the complex relationship between transport and entanglement.
	\item The overall trends indicate that EWCS shows a negative relation with the charge-related conductivity $\sigma_\text{DC}$ and $\alpha_{\text{DC}}$, while HEE parallels the thermoelectric conductivity $\alpha_\text{DC}$. This implicates EWCS is negatively related to charged degrees of freedom, while HEE reflects both charge and thermal excitations.
\end{itemize}

These findings elucidate the essential properties of the Horndeski holographic model and demonstrate the value of entanglement measures in decoding transport phenomena. The contrasting behavior of HEE and EWCS with respect to various parameters emphasizes the need for multiple diagnostic tools to fully characterize quantum information. Moreover, our findings reveal that the dependence of these quantities on the system parameters remains unchanged when comparing scenarios with and without modification terms. This suggests that the observed behavior is a general characteristic of axions minimally coupled to gravity.

While this work focuses on a specific model, the methodology of correlating entanglement and transport properties can offer broader physical insights. Extending the analysis to other holographic models could reveal universal patterns tied to phenomena like momentum relaxation, metal-insulator transitions, and electron correlations. An interesting extension will be to the modification that the axions are not minimally coupled to the Horndenski gravity \cite{Wang:2019jyw}. We expect that this non-minimal coupling can significantly impact the relationship between the entanglement and transport properties. Moreover, modifications like higher derivative terms are also an important topic to further study. However, this could bring in intricate numerical challenges such as the calcultion of fourth-order differential equations. We are working on these directions and will report our findings in future publications.

\section*{Acknowledgments}

Peng Liu would like to thank Yun-Ha Zha and Yi-Er Liu for kind encouragement during this work. This work is supported by the Natural Science Foundation of China under Grant No. 12375048, 12005077, 11905083 and 11805083, as well as the Science and Technology Planning Project of Guangzhou (202201010655) and Guangdong Basic and Applied Basic Research Foundation (2021A1515012374).

\end{document}